\documentclass[sigconf]{aamas}  

\usepackage{booktabs}
\usepackage{amsmath, amssymb}
\usepackage{graphicx,subfigure}
\usepackage{dblfloatfix}
\usepackage{wrapfig}
\usepackage{array}
\usepackage{xr}
\usepackage{xcolor}
\usepackage{algorithm, algpseudocode}

\DeclareMathOperator*{\argmax}{arg\,max}

\makeatletter
\def\BState{\State\hskip-\ALG@thistlm}
\makeatother

\begin{document}


\title{Learning through probing: a decentralized reinforcement learning architecture for social dilemmas }  
\author{Nicolas Anastassacos}
\affiliation{\institution{The Alan Turing Institute} \institution{University College London}}

\author{Mirco Musolesi}
\affiliation{\institution{The Alan Turing Institute} \institution{University College London}}


\begin{abstract}

Multi-agent reinforcement learning has received significant interest in recent years notably due to the advancements made in deep reinforcement learning which have allowed for the developments of new architectures and learning algorithms. However, while they have been successful at solving stationary games, there has been less development in cooperation-type games due to the nature of these algorithms to optimize their play against the opponent's \textit{current} strategy and don't consider how that strategy can change. Using social dilemmas, notably the Iterated Prisoner's Dilemma (IPD) as the training ground, we present a novel learning architecture, \textit{Learning through Probing} (LTP), where Q-learning agents utilize a probing mechanism to determine how an opponent's strategy changes when an agent takes an action. We use distinct training phases and adjust rewards according to the overall outcome of the experiences accounting for changes to the opponents behavior. We introduce a parameter $\eta$ to determine the significance of these future changes to opponent behavior. When applied to the IPD, LTP agents demonstrate that they can learn to cooperate with each other, achieving higher average cumulative rewards than other reinforcement learning methods while also maintaining good performance in playing against static agents that are present in Axelrod tournaments. We compare this method with traditional reinforcement learning algorithms and agent-tracking techniques to highlight key differences and potential applications. We also draw attention to the differences between solving games and studying behaviour using societal-like interactions and analyze the training of Q-learning agents in makeshift \textit{societies}. This is to emphasize how cooperation may emerge in societies and demonstrate this using environments where interactions with opponents are determined through a random encounter format of the IPD.

\end{abstract}

%

\keywords{Reinforcement Learning, Cooperation, Social Dilemmas, Multi-Agent Learning}  

\maketitle


\section{Introduction}


Multi-agent reinforcement learning (RL) has garnered a significant amount of interest in recent years also due to the advancements in deep RL which has allowed for extensive study on agent behaviors. There has been emphasis on designing cooperative agents for decades \cite{Tan1993,Kepetanakis2002} yet extending this success to multi-agent environments has proven difficult as the Markov property is not satisfied since agent behaviors are continuously changing \cite{Sutton1998} and the use of experience replay does little to inhibit unstable learning in presence of multiple learners. Indeed, there are still challenges to be tackled in order to enable broader applications, e.g., in automated decision-making such as self-driving cars, personalized assistants, and the eventuality of artificial agents operating in society. A central aspect of this evolution lies in understanding the competitive and collaborative nature of environments and the emergence of such behaviors \cite{Axelrod1981,Nowak2006}. \\

Humans have cooperated and maintained cooperation to great effect, which has been paramount for the development of civilization \cite{Axelrod1981,Boyd2009,Fehr2004}. Many plants and animals have also demonstrated the tendency to cooperate with relatives and have even been observed cooperating with members of a different species even in highly competitive environments, likely to take advantage of long-term rewards \cite{Clutton2009,Stevens2004}. The evolution of cooperation in competitive environments has therefore been relevant to studies in economics, game-theory, psychology, social science, and now computer science as the future will certainly demand interaction between artificial agents in human and artificial societies. The emergence of cooperative and competitive strategies has been studied in Social Dilemmas \cite{ElinorOstrum1994,ChaoYu2015,VanLange2013,Leibo2017}. These are games where an individual profits from selfishness unless everyone chooses to behave selfishly, in which case the group as a whole achieves an undesirable outcome. In other words, problems arise when too many group members choose to pursue individual profit and immediate satisfaction rather than behave in the group's best long-term interests. From a game-theoretic perspective, the dominant strategy in social dilemmas is often to behave selfishly, which results in arriving at a Nash Equilibria that can be described as \textit{socially deficient} \cite{Leibo2017} and is an undesirable result. One of the first and most studied examples of social dilemmas is the Prisoner's Dilemma (PD), a two-player social dilemma that is formalized by a payoff matrix and a dominant strategy to defect despite a collaborative effort from both players leading to a higher reward. The Iterated Prisoner's Dilemma (IPD) is an extended, sequential version of the PD that has often been the focus for multi-agent RL. In order to succeed in these types of games, it is important for each agent to distinguish between how to play against the current strategy of their opponent and how their opponent's strategy might change to be either more or less cooperative as a result of their own actions as time goes on. \\


Most RL algorithms are designed for single-agent case scenarios. Qualitatively, the Q-value describes how much reward an agent is expected to receive when taking a particular action at a particular state \cite{Sutton1998}. As the environment changes, the learned Q-values become increasingly irrelevant and this is exemplified as the number of agents who are learning increases. Other research projects have aimed to address the issue of non-stationarity in multi-agent environments in a variety of ways ranging from refreshing the experience replay buffer \cite{Leibo2017} using importance sampling \cite{Foerster2017,Uchibe2004} to stabilize learning, using defined policy-types \cite{Lerer2017}, agent-tracking techniques to predict the policy of an opposing agent \cite{Tesauro2004,Foerster2017,Lowe2017,Zhang2010}, and centralized functions to share information across to all participating agents \cite{Lowe2017}. To tackle this, we instead propose a training mechanism, \textit{Learning through Probing}, which allows agents to gather experiences that have been adjusted to reflect behavioral changes in a sequence of events over a period of time via an adjusted reward signal and, therefore, enables them to learn cooperative strategies. Experimentally, we demonstrate that two agents trained with this approach learn to cooperate in the IPD as each agent accounts for the opposing agent's learning while also revealing how their own behavior will change as a result of an opposing agent's chosen actions. Furthermore, we also demonstrate how this type of training mechanism results in a RL agent learning optimal policies for the IPD when matched with other stationary and quasi-stationary strategies from Axelrod tournaments. Finally, we contrast this with current methodologies in multi-agent RL to highlight potential difficulties and we discuss how probing and using experiences through updates might help established methods achieve better performance in dynamic environments. \\


Alongside this architecture, we also demonstrate that adjusting the training environment can lead to cooperative behaviors using a standard Q-learning algorithm to take into account the effect of external factors beyond each individual agent's decision-making in determining what behaviors can emerge when introducing artificial agents into open environments like societies. The focus in multi-agent RL is predominantly based around stationarizing the environment in order for agents to learn how to achieve optimal outcomes in closed environments. This typically requires the same agents to be used in training and testing. However, an overlooked aspect of agent behavior is how to design environments to \textit{nurture} certain types of behavior. We present experimental results on how untrained agents can learn to cooperate with other untrained agents using standard Q-learning when interacting with static agents at the same time. 


\section{Related Work and Motivation}

Modeling multi-agent systems and designing learning algorithms for dynamic settings is hard and the majority of work in this area focuses on competitive, zero-sum games. A common approach is to simply have each agent treat all other agents as part of the environment and learn independently, however, this generally leads to less than optimal performance \cite{Kepetanakis2002,Shoham2003}. Firstly, a naive implementation of experience replay is not suitable for dynamically changing environments \cite{Adam2012,Schaul2015}. Others have employed importance sampling in order to stabilize the application of replay buffers where experiences that are collected from ``old" environments can still be used to update Q-values \cite{Uchibe2004,Foerster2017}. Secondly, agents have to account for actions taken by other agents in their Q-value approximations \cite{Busoniu2010}. Finally, learning to play optimally against just an opponent's \textit{current} behavior may trap players in undesirable states, as can be the case in social dilemmas, as there is no consistent and suitable method of exploration. \\

However, while Q-learning and other reinforcement learning algorithms have shown to perform well in zero-sum games for single player, there has been less work in analyzing their resulting behavior when trained in general-sum games. While in zero-sum games, the rewards between two opposing agents are negatively correlated, in general-sum games, the rewards between two opposing agents can be correlated arbitrarily as the resulting outcome for one individual doesn't necessarily directly impact the outcome for others. This case is generally more representative of interactions in the real-world. The accepted baseline solution for such games is the Nash equilibrium. In a Nash equilibrium, each player has chosen a strategy and neither player can benefit by changing their strategy unilaterally and it is proven to exist for every finite game. Inspired by this approach, several game-theoretic reinforcement learning algorithms have been proposed to solve general-sum games such as Nash Q-learning. This algorithm modifies its update such that its update is based on the expectation that agents would take their equilibrium actions \cite{NashQ2003}. \\

There is a parallel here between the goal of agent-tracking methods and Nash Equilibrium as agent-tracking methods look to identify the opponent's strategy and update assuming only small changes in the opponent's behavior. However, while they may account for the opponent's strategy, they both look to perform updates conditioned on the opponent's current behavior. This is a flawed approach for cooperation games as a key aspect of the game involves adapting to your opponent's behavior and therefore doesn't consider how an agent's learning influences how its opponent's behavior will change. \\

The study of sequential social dilemmas, notably the IPD, has been prevalent across numerous disciplines such as game theory, economics, and across social sciences as a way of analyzing complex behaviors such as altruism, reciprocity and cooperation \cite{Rapoport1974,Macy2002,Nowak2005}. A strategy known as Tit-For-Tat (TFT) where an agent cooperates on the first move and then replicates an opponent's previous action, known as \textit{equivalent retaliation}, is a simple strategy yet has shown to be one of the most effective strategies in the IPD and has served as a basis for the modeling of many real-world behaviors \cite{Axelrod1981}. Social dilemmas like the IPD have proven to be an effective training ground for multi-agent RL as they might involve cooperation as a viable method of achieving optimal performance against fixed policies as well as learning agents \cite{Sandholm1996,Zhang2010,Lerer2017}. In order to perform well in social dilemmas, agents must learn to forgo their desire for early rewards and ``agree" on strategies that will benefit those involved. Many agents designed for Axelrod tournaments utilize reciprocating behavior which is the general philosophy behind the Tit-For-Tat strategy whereas others employ more grudging techniques to manoeuvre opponents into cooperating. \\

Leibo et al. have attempted to train Q-learning agents for sequential social dilemmas so as to analyze how conflict, competition, and cooperation can emerge via a multiplayer Wolfpack hunting game and a Gathering game \cite{Leibo2017}. Replay buffers of a fixed capacity were used to try and accommodate for the multiple learners; once filled, they were refreshed so that the agent emphasized training on more recent experiences. A more recently popular approach to multi-agent reinforcement learning involves \textit{policy prediction} which, alongside Q-learning (independent learners), we will contrast our approach with. We emphasize this approach because we think that it closely captures a necessary element of multi-agent learning: understanding opponent behavior and incorporating it directly into learning. It expands on methods like fictitious play \cite{Erev1998} and joint action learning (JAL) to more accurately represent an opponent's policy \cite{Claus1998,Marden2009} and enables coordinations (in cooperative games). Tesauro presents Hyper-Q learning, which learns the value of mixed strategies instead of base actions by estimating opponent actions using observed data and evaluates it using Rock-Paper-Scissors \cite{Tesauro2004}. He further argues that  Hyper-Q learning may be effective against agents even if they are persistently dynamic. This has been corroborated by other research employing similar philosophy of policy prediction. Experiments on Starcraft and other abstract games that require complex multi-agent coordination have shown that this methodology significantly improves performance compared to independent learners trained with Q-learning though they are often combined with other techniques, e.g., sampling techniques and a centralized value function \cite{Foerster2017,Lowe2017}. However, as we will demonstrate in our experiments, these types of methods do not perform adequately in social dilemmas as they aim to shape learning around what is happening currently rather than what \textit{could} happen in the future. In contrast, our approach focuses directly on understanding the consequence of actions on opponent's behavior and incorporates that knowledge directly into agent learning via an adjusted reward function. \\
The use of social behavior metrics is another approach to tackle the issue of describing what is really happening in a state at any moment in time in decentralized learning environments \cite{Perolat2017}. However, it is difficult to determine how these metrics should be designed as they are contextually dependent on the environment. Matignon et al. achieve cooperative behavior in a decentralized RL system using a modified update equation that is conditioned on the size of the reward \cite{Matignon2007}. Another decentralized method by Yu et al. attempts to embed emotional context into agents to drive them to learn cooperative behaviors using various metrics to represent an agent's drive and emotions relative to neighbouring agents \cite{ChaoYu2015}. However, these approaches represent only the current standings that are available without an indication of how things may or may not change which we maintain is essential to developing cooperative behavior. An approach that also emphasizes integrating future behavior of one's opponents is LOLA which looks to consider opponent learning, optimizing its return using a one-step look ahead which requires direct access to the opposing agent's parameters \cite{Foerster2018}. Our approach differs in a number of ways. Firstly, we identify two distinct phases. The first phase is the \textit{probing phase} where each agent can probe the opponent agent in order to gather information about how their opponent's strategy changes after an update and adjust any collected experiences. We use a defined time horizon to determine the number of updates to the opponent's strategy to consider. In the second phase, the agent trains on the adjusted experiences only. Secondly, with the addition of the probing phase, the agent's do not need to have information about the parameters of the opponent agent or need to track their strategy in advance. \\

\section{Preliminaries}

\subsection{Q-Learning}
Q-learning is a popular off-policy reinforcement learning technique to learn optimal behaviors. An agent trained with Q-learning looks to take actions that maximize its expected cumulative reward. A value function for a policy $\pi$ is given by 
\begin{equation}
    V^\pi(s) = \mathbb{E}[ \sum_{i=1}^{T} \gamma^{i-1}r_i | s_0 = s, \pi] \quad \forall s \in \mathbb{S}
\end{equation}

Among all possible value functions there exists a maximum optimal value function $V^* = \max_{\pi} V^\pi(s) \quad \forall s \in \mathbb{S}$ and an optimal policy that corresponds to the optimal value function $\pi^{*} = \argmax_{\pi} V^\pi(s) \quad \forall s \in \mathbb{S}$. The  Q-function, $Q$, is defined as the expected cumulative reward received by an agent starting in $s$, picking action $a$ and behaving optimally from that point onward. We can therefore write the optimal Q-function as
\begin{equation}
Q^*(s,a) = r(s, a) + \gamma \mathbb{E}_{p(s'|s ,a)}[V^*(s')]
\label{optimal-q}
\end{equation}
From our definition of the optimal value function, we can derive that $V^{*} = \max_{a} Q^*(s, a)$ and therefore, $\pi^{*} = \argmax_{\pi} Q^\pi(s, a)$. The optimal policy is therefore the policy that chooses the action with the highest Q-value at every state. However, we can see clearly in \eqref{optimal-q} that the optimal Q-value is subject to state transition distribution remaining the same. This is sensible in single-agent games but not in multi-agent settings since we expect the state distribution to change as agents change their behavior and therefore affect the resulting state transitions as they are learning. Agent-tracking methods expand on this, conditioning the transition to $s'$ on both the agent's own action as well as any opponent's actions in order to better approximate their Q-values, however, as we will see, they fail to achieve the desired solution for cooperation-type games like the IPD.

\subsection{Hyper-Q-Learning}

Hyper-Q-learning is an agent-tracking technique and an extension of Q-learning for multi-agent systems. It estimates an opponent's mixed strategy $y$ and then evaluates the best response. In the single agent case, the Hyper-Q function $Q^\pi(s, y, a)$ is adjusted such that 
$$ Q^\pi(s, y, a) = r(s, y, a) + \gamma \max_{a'} Q^\pi(s', y', a') $$ 
where $y'$ is a new estimated opponent strategy in $s'$. Variations of Hyper-Q have performed well using deep neural networks with sampling modifications to the replay buffer. A similar approach in multi-agent scenarios has been implemented with success using actor critic techniques such that the target value of agent $j$ is $$ \hspace{-5.4cm} Q^{\pi_j}(s,a_1,...,a_N) = $$ $$ r_j(s, a_1,..., a_N) + \gamma \mathbb{E}_{s'} \max_{a'}Q^{\pi_j}(s', a'_1,...,a'_N) |_{a_{-j}=\pi_{-j}} $$ 
where $-j$ denotes all agents except of agent $j$. In this paper, we use a variation of Hyper-Q for simplicity in the IPD adopting a separate neural network to estimate an opponent's next action directly from an observation and optimize by taking recent samples of an agent from a replay buffer. While agent-tracking methods have performed better than independent learners, they still suffer from a problem of non-stationarity, which is that the distribution of states changes as each agent updates their policy . If the updates to $\pi_{-j}$ are small and the environment is less dynamic then $p_{i}(s'|s, a_1,...,a_N)|_{a_{-j}=\pi_{-j}} \approx p_{i+1}(s' | s, a_1,..., a_N)|_{a_{-j}=\pi_{-j}}$, and is sufficient for calculating an optimal policy though this problem is made more difficult with more learners. Furthermore, since agents cannot anticipate how behavior will change in the future, they cannot avoid getting trapped in socially deficient Nash Equilibria (which are optimal and dominant strategies given knowledge only of an opponent's current policy). We contrast this with our approach that, instead, emphasizes an awareness of how an agent's actions can influence an opponent's future response as we argue that optimizing against the current policy of an opponent can trap agents into policies that result in both parties receiving inadequate rewards.

\subsection{Forms of Iterated Prisoner's Dilemma}
The Prisoner's Dilemma (PD) is a simple game that serves as the basis for research on social dilemmas. The premise of the game is that two partners in crime are imprisoned separately and each are offered leniency if they provide evidence against the other. Each player can choose between two actions: cooperation (C) or defection (D), and the payoffs of the game are displayed in Figure \ref{figure: social dilemma}. The game is modeled so that $T > R > P > S$ and $2R > T + S$. Solving this from a game-theoretic perspective, the dominant strategy is to defect, however, if both players take this action then they arrive at a Nash Equilibrium that is socially deficient. Originally, the PD is a one round game, but the IPD is a \textit{sequential} PD often studied to understand the effects of previous outcomes and the emergence of cooperative behaviors.

\begin{figure}[!ht]
\centering
\begin{tabular}{c | c | c}
 SD & C & D \\
 \hline
 C & \textit{R, R} & \textit{S, T} \\
 \hline
 D & \textit{T, S} & \textit{P, P} \\
\hline
\end{tabular}
\quad \quad \quad
\begin{tabular}{c | c | c}

 IPD & C & D \\
 \hline
 C & 3, 3 & 0, 5 \\
 \hline
 D & 5, 0 & 1, 1 \\
\hline
\end{tabular}

\caption{Payoff Matrix for Social Dilemmas and Iterated Prisoner's Dilemma. The motivation to defect comes from fear of an opponent defecting or acting greedily to gain the maximum reward when one anticipates the opponent might cooperate. }
\label{figure: social dilemma}
\end{figure}

\section{Learning through Probing}

In this section, we present and summarize the architecture and learning methodology of the probing technique that we apply to RL agents. Each agent consists of two separate components that we term the \textit{probe} and the \textit{player}. Each player $i$ has a policy $\pi_{p_i}$ parameterized by $\theta_{p_i}$. Each \textit{probe}, similarly, has a policy $\pi_{b_i}$ parameterized by $\theta_{b_i}$. The role of the probe is to generate experiences $(s_t, a_t, r_{t+1}, s_{t+1})$ that account for opponent strategy changing due to learning when performing action $a_t$ after observing $s_t$. The eventual consequences of this action are measured by aggregating the rewards of the total sequence of events $\tau$ over a finite time horizon $T$. This is an adjusted experience to capture the affects of taking that action on the opponent's policy which will eventually be used to train the player component. To start, each probe must gauge the consequences of a type of action. Paired with an opponent, the probes explore the environment and store experiences in a replay buffer. These experiences are then grouped according to the actions of the opponent. The probe then updates on the subset of experiences and continues to play versus its opponent. Each \textit{probe} update is based on the set of \textit{opposing} experiences stored. After taking a one-step update based on these initial experiences, the \textit{probes} play against each other according to their learned policies and repeat the process for $T$ updates. Eventually, the sequence $\tau = (s_0, a_0, r_1, s_1,..., s_{T-1}, a_{T-1}, r_{T}, s_{T})$ is stored. \\

During the probing phase, the experiences are grouped by the tuple $(a_1, a_2)$, the agent's action and opponent's action similarly to JAL and agent-tracking methods. Alternatively, actions could grouped according to the reward outcome. In the context of the PD, grouping actions according to the opponent's action or the reward outcome is the same, however, in general, grouping while considering the opponent's action is less ambiguous. \\

After the probes have interacted and generated experiences, the rewards of the resulting sequences $\tau$ are adjusted such that $R(\tau) = \sum^T_{t=1} \eta^{t-1} r_t$ and are used to train the \textit{player} components of the agents. $\eta$ is an added \textit{discount through updates} term to determine how many future interactions the agent should consider. By manipulating the value of $\eta$ we can determine how each \textit{player} values the approximated long-term outcome and $\eta = 0$ indicates an approach identical to Q-learning. The gradient of the \textit{player} updates according to 

$$ Q^{\pi_{i+1}}(s, a) \leftarrow Q^{\pi_{i}}(s, a) - R(\tau) + \gamma \max_a Q^{\pi_{i}}(s', a') $$ 

After training has concluded on the adjusted experiences, the learned policy is transferred to the probe. When only one of the participants is learning to maximize an RL objective function the architecture is adjusted so that the \textit{probe} interacts directly with the opponent rather than establishing a probing phase for both agents. \\

\begin{figure*}
	\centering
	\includegraphics[width=0.615\textwidth]{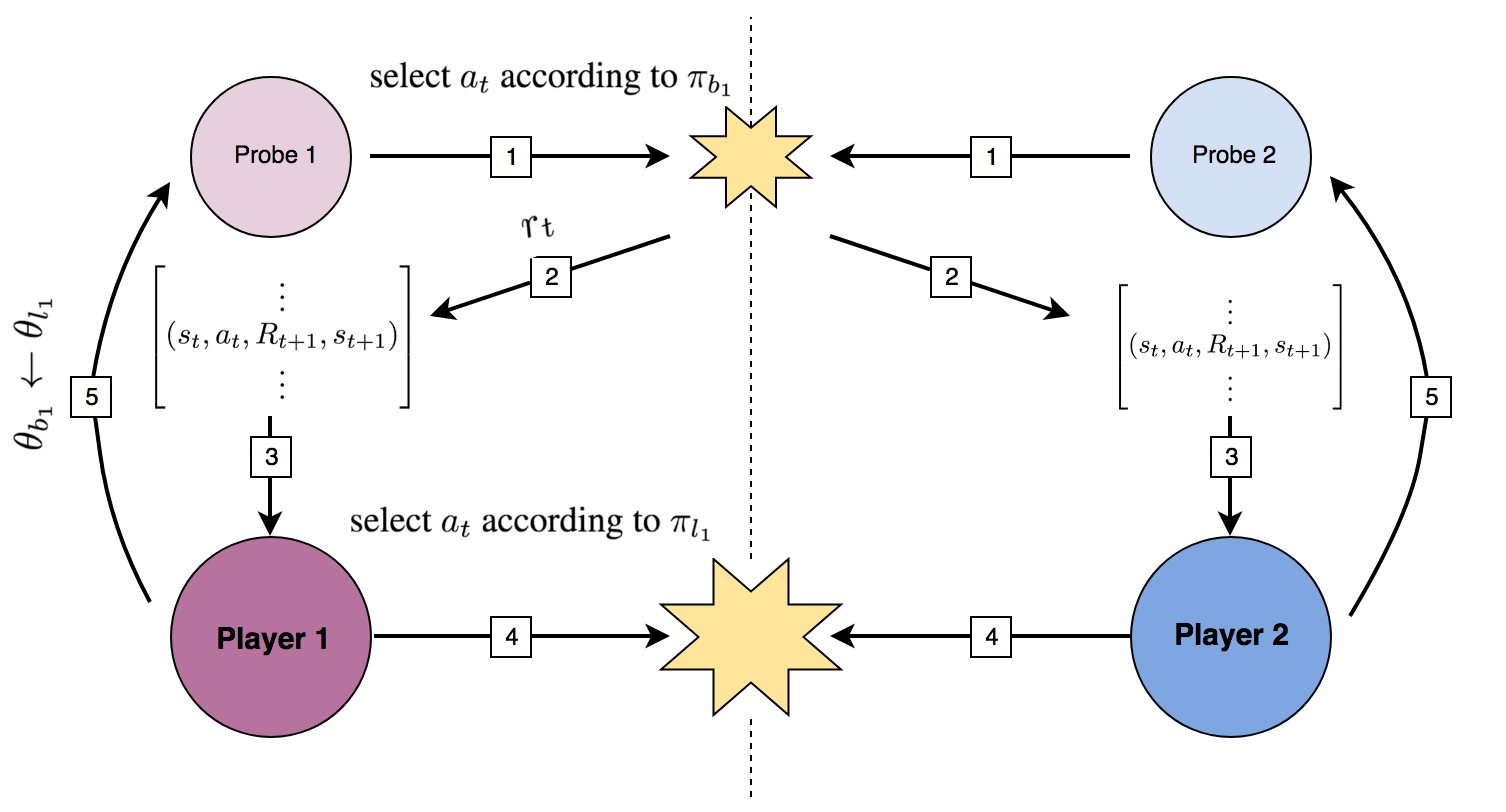}
    \label{ltp_arch}
    \caption{Learning through Probing architecture diagram involving two RL agents. 1) After exploring the environment, the \textit{probe} component trains on subsets of experiences to learn consequences for actions. Actions are then selected according to a learned policy. 2) Experiences are collected into a replay buffer and adjusted. 3) The \textit{player} component trains on the adjusted experiences. 4) The players are matched against each other after training. 5) In continuously adaptive games, probes could adopt learned player policies and adapt their strategies over time.}
\end{figure*}

\begin{algorithm}
    \caption{Learning through Probing}\label{euclid}
    \begin{algorithmic}[1]
    
    \State $\textbf{Input:} \gets \theta_i^{b_1}, \theta_i^{b_2} $
    \For { i \textbf{in} $range(0, \text{episode})$}
        \State $o^{b_1}_i, \quad o^{b_2}_i \gets Dilemma(agent_1, agent_2)$
        \State $a^1_i \gets \theta^{b_1}_i, \quad a^2_i \gets \theta^{b_2}_i$
        \State $(o^{b_1}_{i+1}, r_i^{b_1}), (o^{b_2}_{i+1}, r_i^{b_2}) \gets Dilemma.play(a^1_i, a^2_i)$
        \State $\mathit{replay}\text{\textunderscore} \mathit{buffer}_1.append(o^{b_1}_i, a^1_i, r^1_i, o^{b_1}_{i+1})$
        \State $\mathit{replay}\text{\textunderscore} \mathit{buffer}_2.append(o^{b_2}_i, a^2_i, r^2_i, o^{b_2}_{i+1})$
    \EndFor
    \State $\text{group experiences according to joint actions} (a^1_i, a^2_i)$
    \State $\text{ and store in separate buffers}$
    \For {$\mathit{buffer} \text{\textbf{ in }} \mathit{replay\text{\textunderscore}buffers}$}
    	\State $\text{Update } \theta_t^{b_1}, \theta_t^{b_2}$
        \State $o^{b_1}_t, \quad o^{b_2}_t \gets buffer$
        \For {$t \textbf{ in } range(0, T)$}
    	    \State $a^1_t \gets \theta_t^{b_1}, \quad a^2_t \gets \theta_t^{b_2}$
    	    \State $(o^{b_1}_{t+1}, r_i^{b_1}), (o^{b_2}_{t+1}, r_t^{b_2}) \gets Dilemma.play(a^1_t, a^2_t)$
    	    \State $\text{Update } \theta_t^{b_1}, \theta_i^{b_2}$
    	\EndFor
    	\State $\tau_1 \gets (o^{b_1}_t, a^1_t, r^1_t, o^{b_1}_{t+1}, a^1_{t+1}, r^1_{t+1},..., a^1_T, r^1_T, o^{b_1}_{T+1})$
    	\State $\tau_2 \gets (o^{b_2}_t, a^2_t, r^2_t, o^{b_2}_{t+1}, a^2_{t+1}, r^2_{t+1},..., a^2_T, r^2_T, o^{b_2}_{T+1})$
    \EndFor
    
    \State $\text{reset } \theta^{b_1}_i, \theta_i^{b_2}$
    \State Adjust overall sequences $\tau_1, \tau_2$
    \State Train players, $\theta^{p_1}_i, \theta^{p_2}_i$ on adjusted experiences, $\tau_1, \tau_2$

    \end{algorithmic}
\end{algorithm}

\section{Implementation and Experiments}

In this section, we describe the methodology and experimental setup. The first subsection describes how RL agents train against Axelrod agents simultaneously, and compares and contrasts the results of Q-learning and Hyper-Q-learning to examine how policy prediction improves learning under certain conditions. The second subsection describes how we tackle these difficulties using a \textit{probing} technique and Q-learning to account for behavior changes \textit{through updates} and investigating the influence of this new learning methodology on the emergence of cooperation. The default neural networks had two hidden layers with 40 hidden units and ReLU activation functions. Agents were trained with gradient descent with a buffer size of 1e5, learning rate of 1e-4 and a batch size of 300. Exploration, $\epsilon$, was initially set to 1.0 and decreased linearly with iterations, stopping at 0.1. The discount rate $\gamma$ was set to 0.99. Target networks were also used to further stabilize training. After testing multiple values for the time horizon, $T$, we set it to 5. Using values greater than 5 produced results that were very similar while using values smaller than 5 had a higher variance. Finally, the third section describes the external aspects of learning in artificial societies and the impact on agent behavior.

\subsection{RL Agents versus Axelrod Agents}
The Axelrod library \cite{AxelrodPython} contains an extensive set of strategies that have been used in previous Axelrod tournaments as well as those that have been rigorously covered in scientific literature. We will refer to the collection of these strategies as ``Axelrod agents''. All strategies in the tournament follow a simple set of rules: players are unaware of the number of turns in a match, players carry no acquired state between matches, players cannot observe the outcome of other matches and players cannot manipulate or inspect their opponent in any way beyond what is required in a match. We highlight a diverse set of strategies taken from the Axelrod library that will be used as opponents. \\

\begin{figure*}[!t]
  \centering
  \subfigure[\textbf{Average cumulative rewards and standard deviation for various RL agents over 20 encounters at numerous training iterations.}]{\includegraphics[scale=0.41]{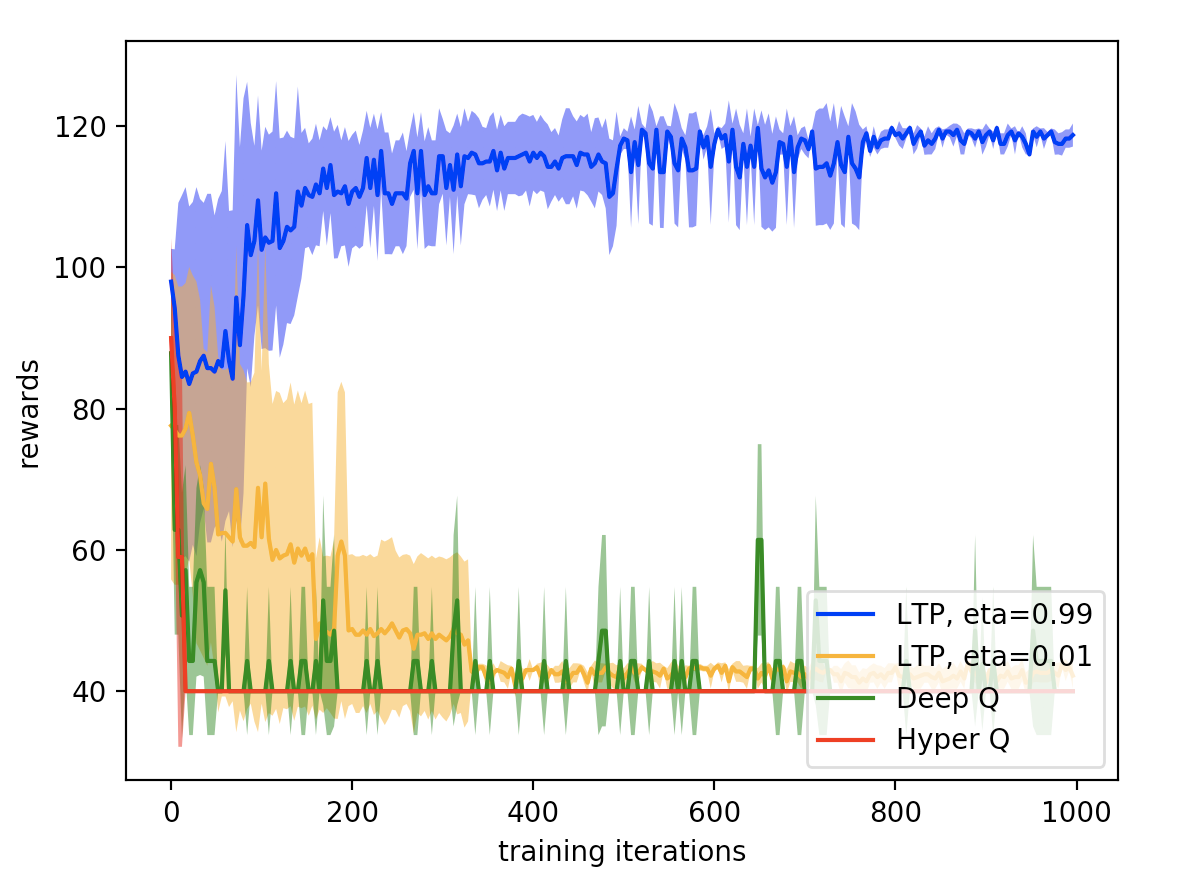}} \quad
  \subfigure[\textbf{Average rewards and standard deviation of LTP agents over 20 timesteps trained with various $\eta$ values.}]{\includegraphics[scale=0.405]{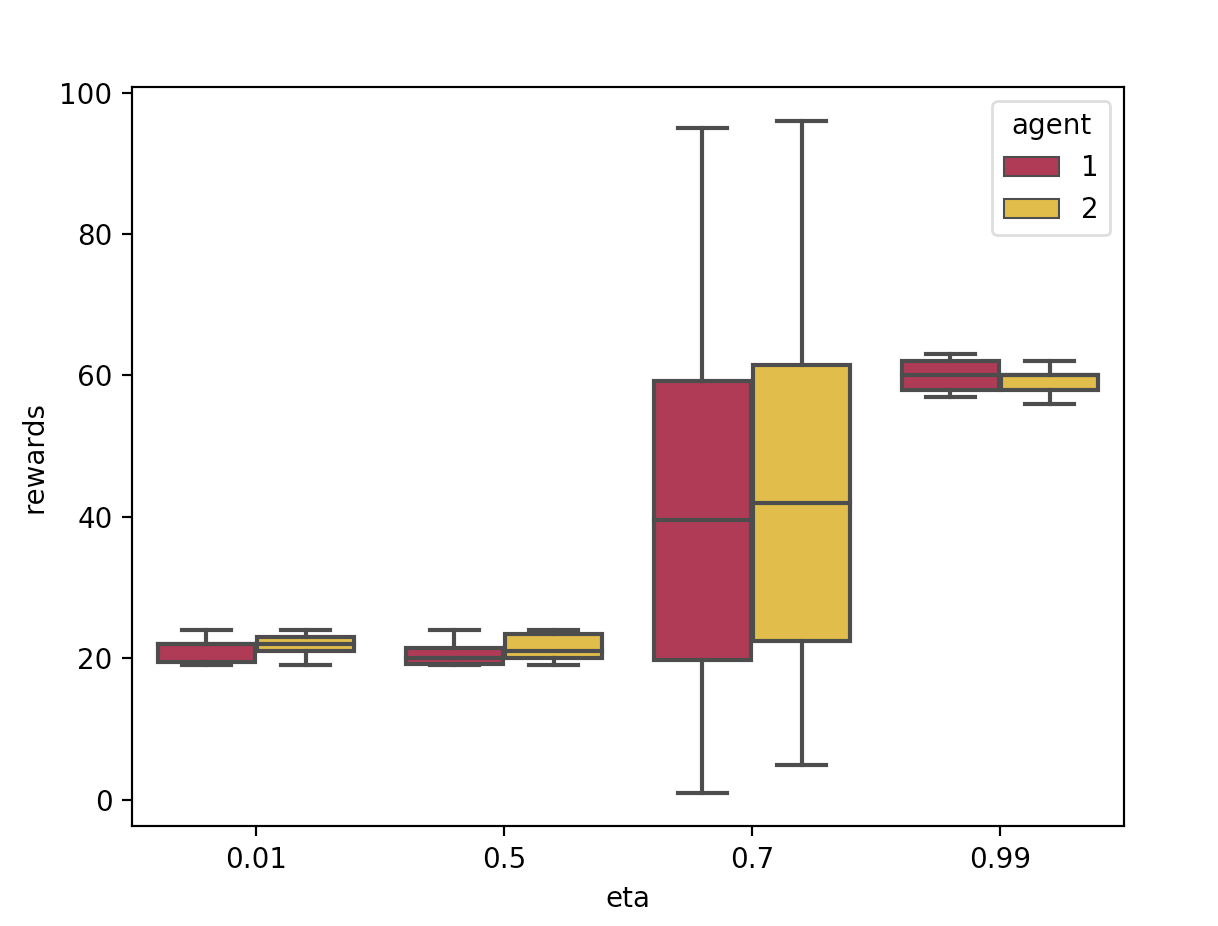}}
  \caption{(a) The blue line shows the results achieved by LTP agents trained with $\eta=0.99$. These agents learn to cooperate early and consistently with decreasing variance with more training iterations indicating stable performance. Trained with $\eta=0.01$, LTP agents are slower to converge than Deep-Q or Hyper-Q agents, however, achieve similar results with consistency. Hyper-Q agents are the quickest in learning to defect. (b) Agents trained with higher $\eta$ values learn cooperative policies with little error rate. A threshold value specific to the used configuration of the IPD is noted at approximately $\eta=0.7$.
  \label{Figure: avg RL rewards}}
\end{figure*}

\textbf{Tit for Tat (TFT).} The Tit for Tat strategy is a forgiving strategy that will cooperate on the first move and then perform the same action as the opponent's most recent move. 

\textbf{Punisher.} The Punisher strategy is a grudging strategy that starts by cooperating, however, if at any point its opponent defects, it will defect for \textit{memory length} where \textit{memory length} is proportional to the opponent's historical percentage of defecting.

\textbf{Forgetful Grudger.} The Forgetful Grudger strategy is a grudging strategy which defects for a fixed length of time if an opponent defects at any point. If an opponent cooperates for long enough it forgets its grudge and will cooperate until it sees another defection.

\textbf{Prober.} The Prober strategy plays an initial sequence of moves to feel out an opponent's strategy. It keeps a count of defections that are retaliating and defections that are unmerited. If the number of justified defections and number of unjustified defections differs by more than 2, cooperate for the next 5 turns and then play TFT's strategy. Otherwise defect forever. 

\textbf{Sneaky.} The Sneaky strategy is an original strategy that tracks the three most recent actions of the opponent. If all three actions are to cooperate, then it will defect. Also, if the total number of opponent defections are greater than the total number of opponent cooperations then it will defect. \\

Some of these agents, like the Prober, are not stationary to begin with but converge to stationary distributions given enough time. We refer to these as quasi-stationary behaviors. Q-learning, Hyper-Q-learning and LTP agents are each trained to play against the above agents. Agents are able to observe the previous four actions taken by themselves and their opponent. Each match lasts for 100 timesteps. Agents are paired using a round robin matchmaking format. This was run for 10,000 episodes. For Q-learning and Hyper-Q-learning agents, $\epsilon$ was set to 1.0 initially and decreased linearly with iterations until it reached 0.1.



\subsection{RL Agents versus RL Agents}
We carry out experiments in order to see how Q-learning agents, Hyper-Q-learning agents and LTP agents perform against their counterparts in the IPD. Each state was characterized by the agent's and its opponent's previous four interactions. Agents had no information about who they were playing with beyond what was available in the observation data. While versus Axelrod agents, though Q-learning and Hyper-Q learning agents keeping an exploration rate at 1.0 is feasible when playing against static strategies, Hyper-Q relies on developing accurate predictions of an opponent's next move and so we ensure that the observed policies are non-random. We train LTP agents against one another using $\eta$ values of 0.99, 0.7, 0.5, and 0.01 to demonstrate the effect of accounting for behavioral adaptations. These agents play against each other for 1000 iterations. After each iteration agents would play the IPD for 20 timesteps and the cumulative reward was recorded.


\begin{figure*}[!h]
  \centering
  \subfigure[0 TFT agents]{\includegraphics[scale=0.11]{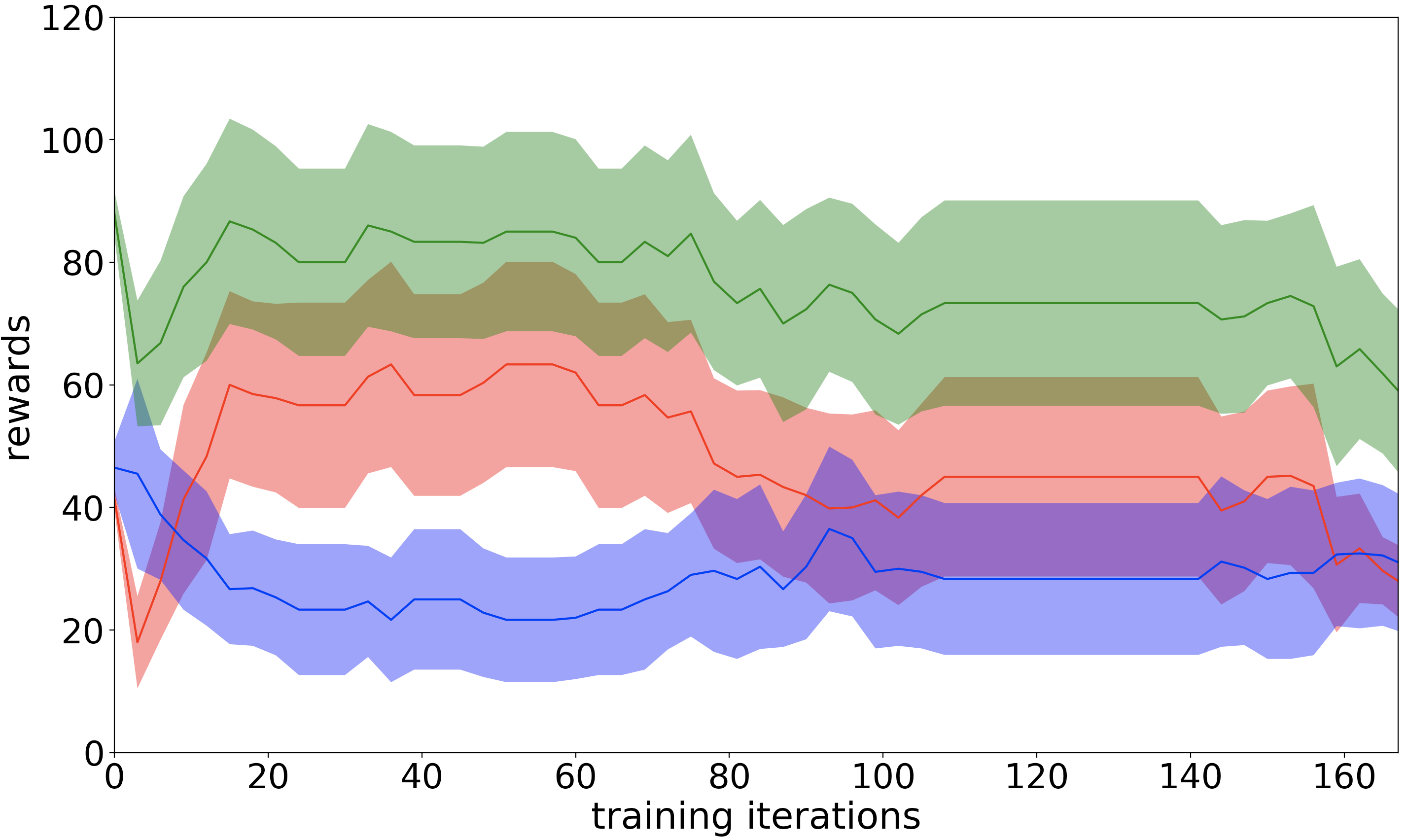}} \quad
  \subfigure[1 TFT agents]{\includegraphics[scale=0.11]{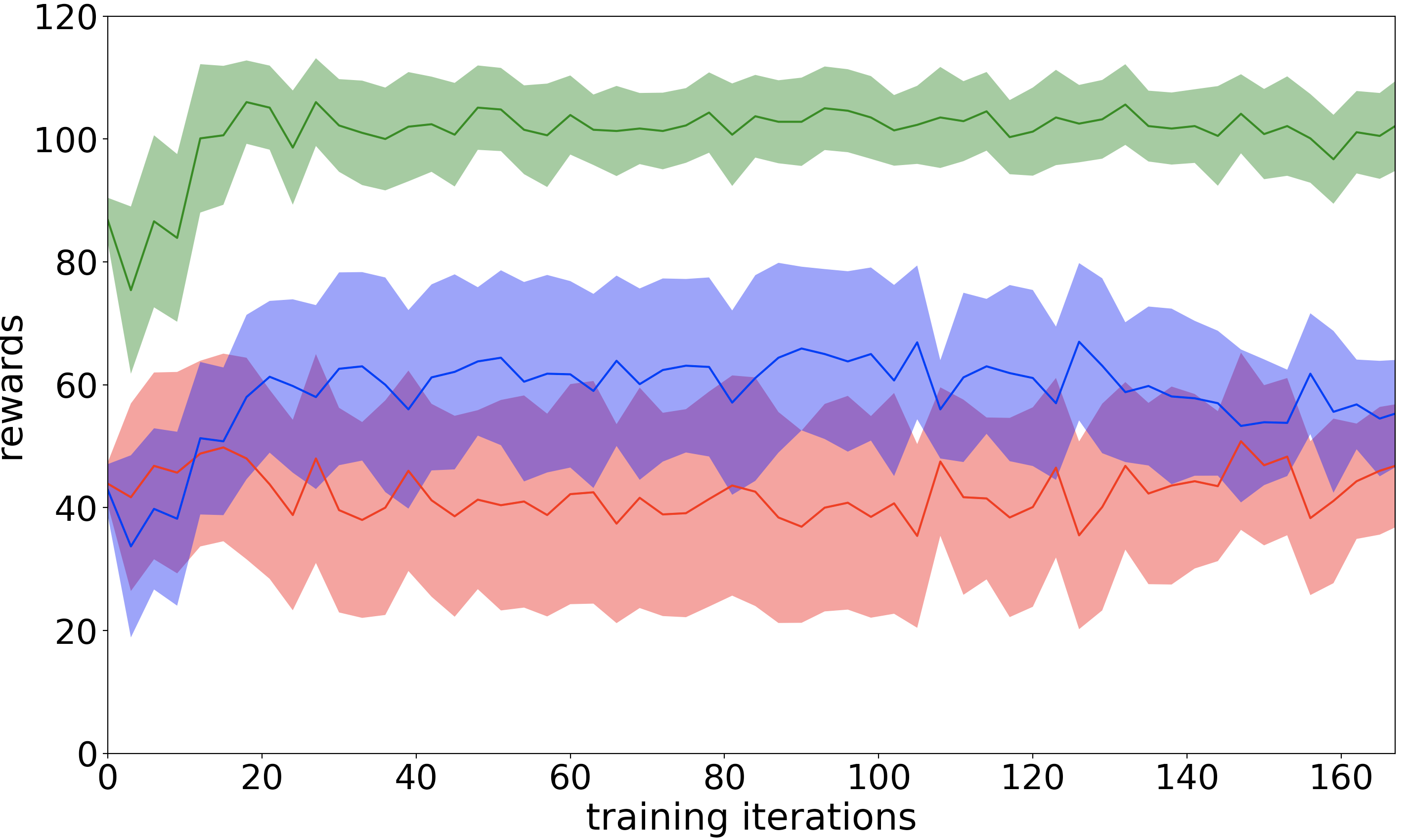}} \quad
  \subfigure[2 TFT agents]{\includegraphics[scale=0.11]{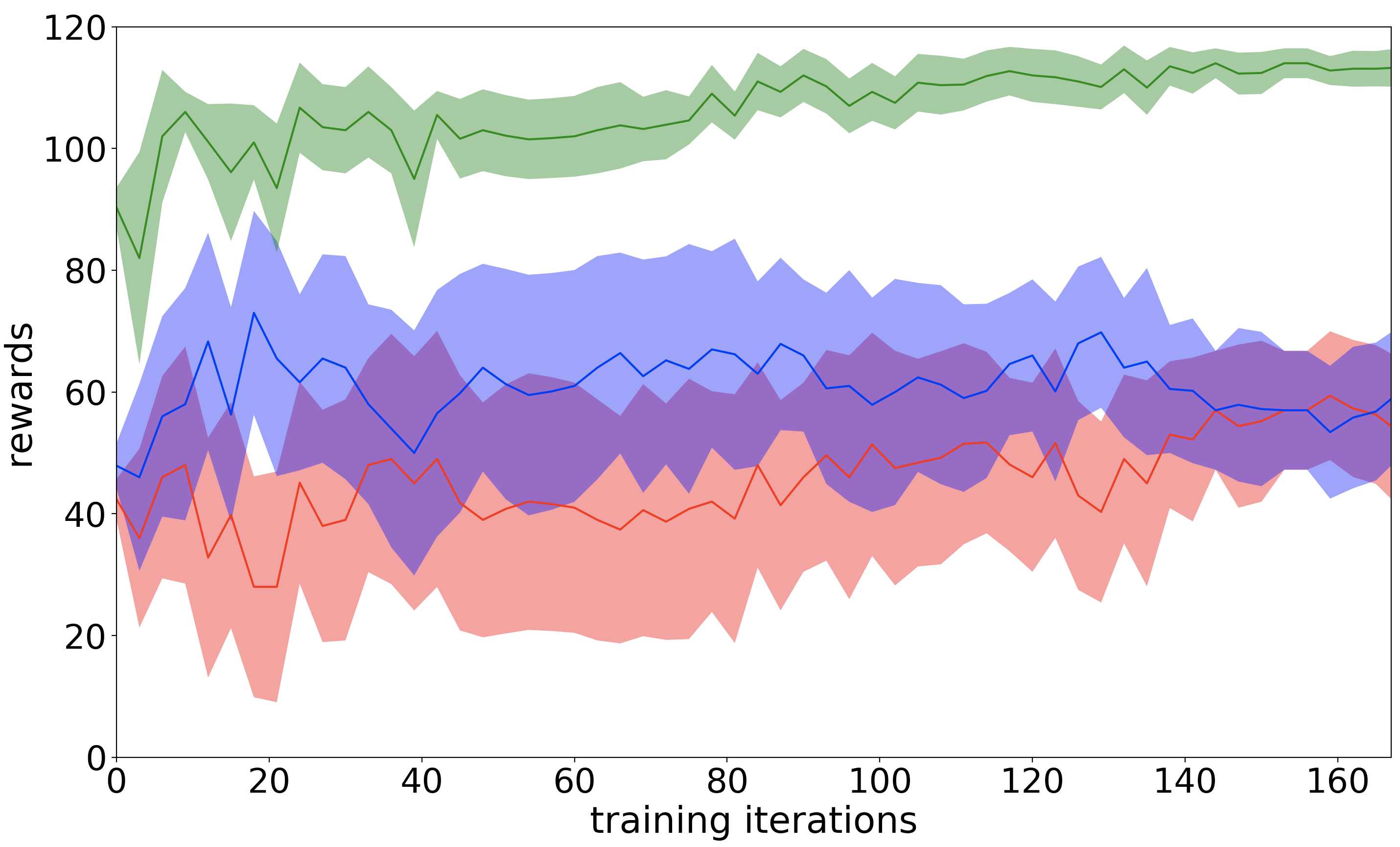}} \quad
  
  \subfigure[1 Sneaky agent, 0 TFT agents]{\includegraphics[scale=0.11]{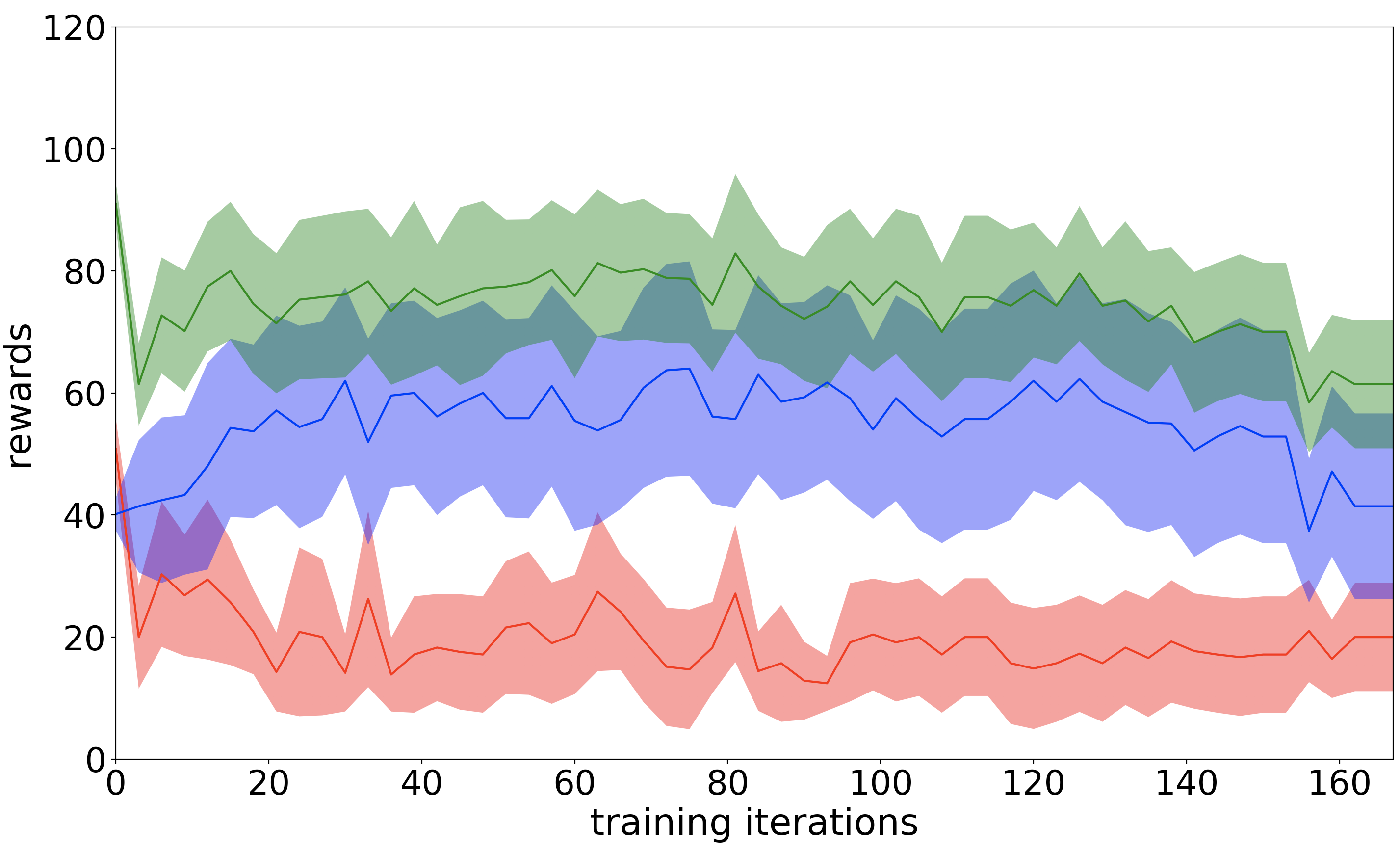}} \quad
  \subfigure[1 Sneaky agent, 1 TFT agents]{\includegraphics[scale=0.11]{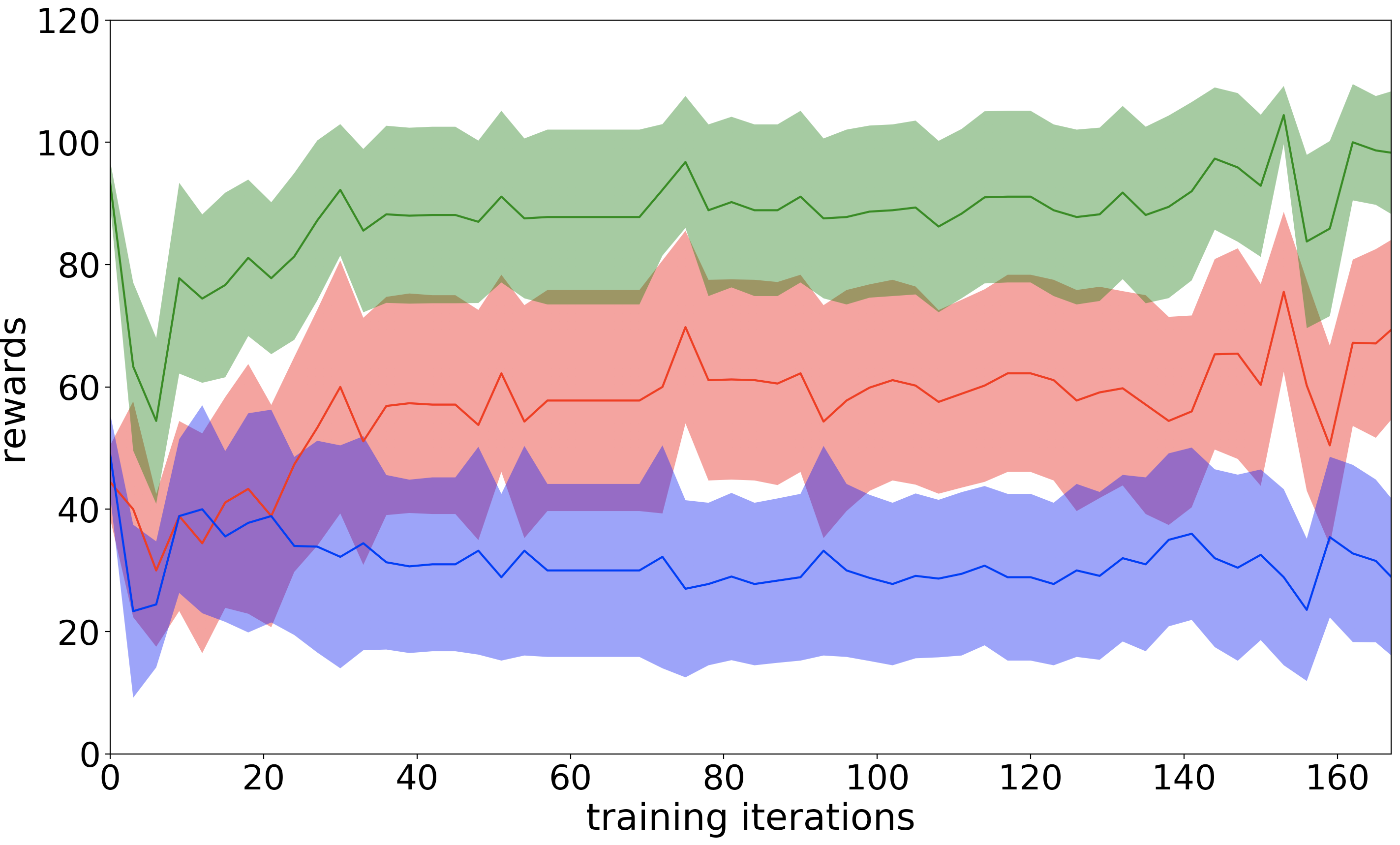}} \quad
  \subfigure[1 Sneaky agent, 2 TFT agents]{\includegraphics[scale=0.11]{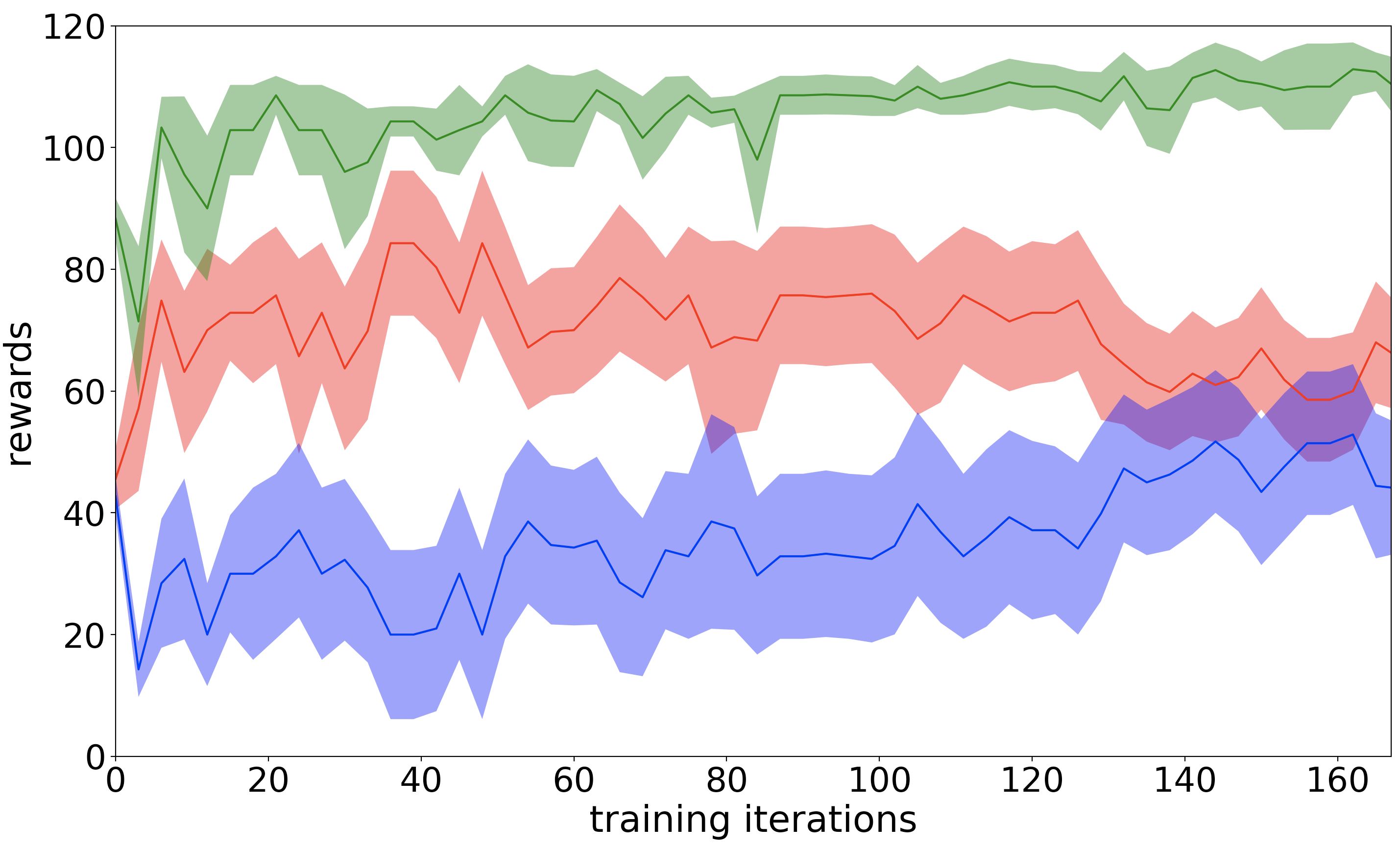}} \quad
  
  \subfigure[1 Punisher agent, 0 TFT agents]{\includegraphics[scale=0.11]{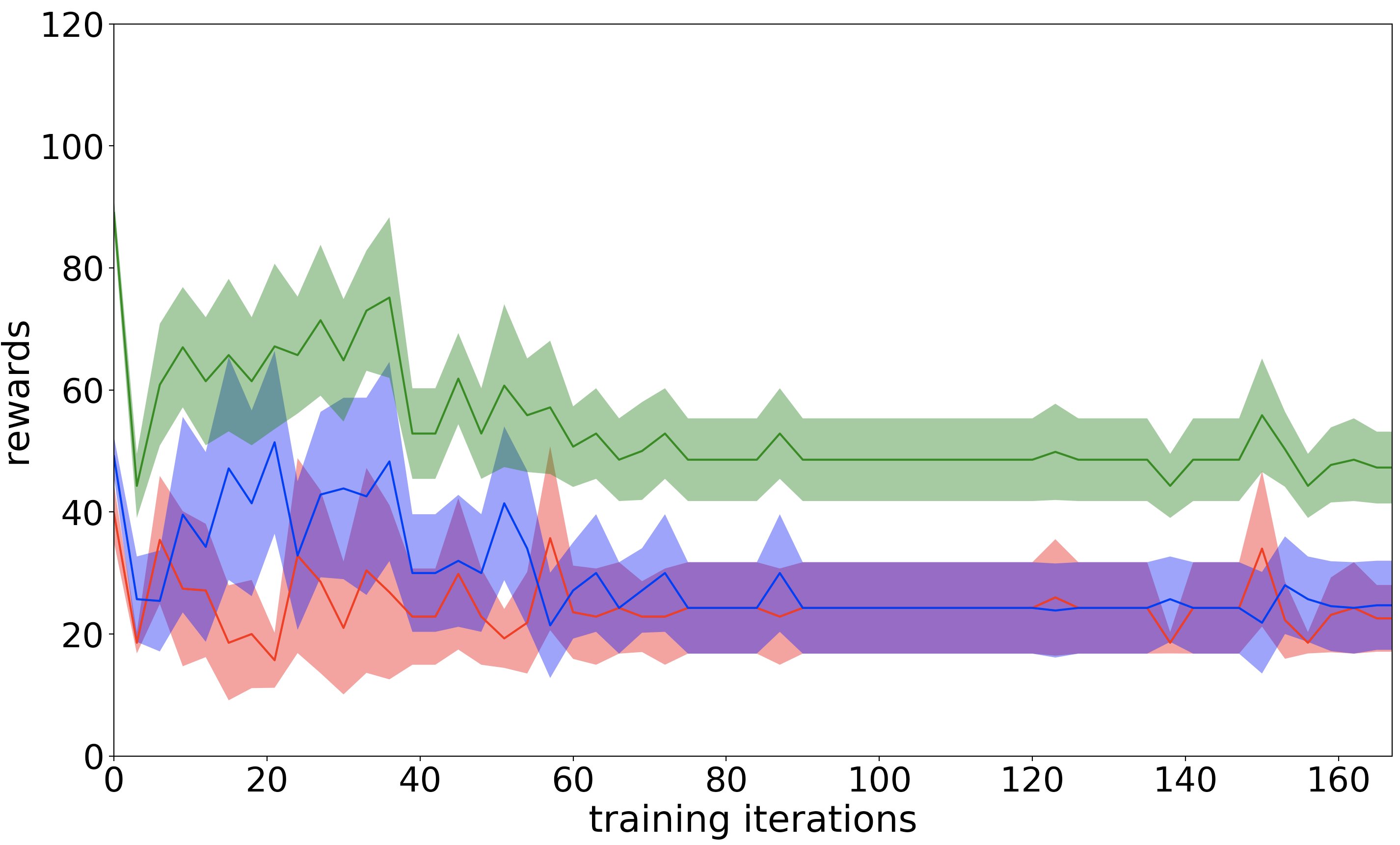}} \quad
  \subfigure[1 Punisher agent, 1 TFT agents]{\includegraphics[scale=0.11]{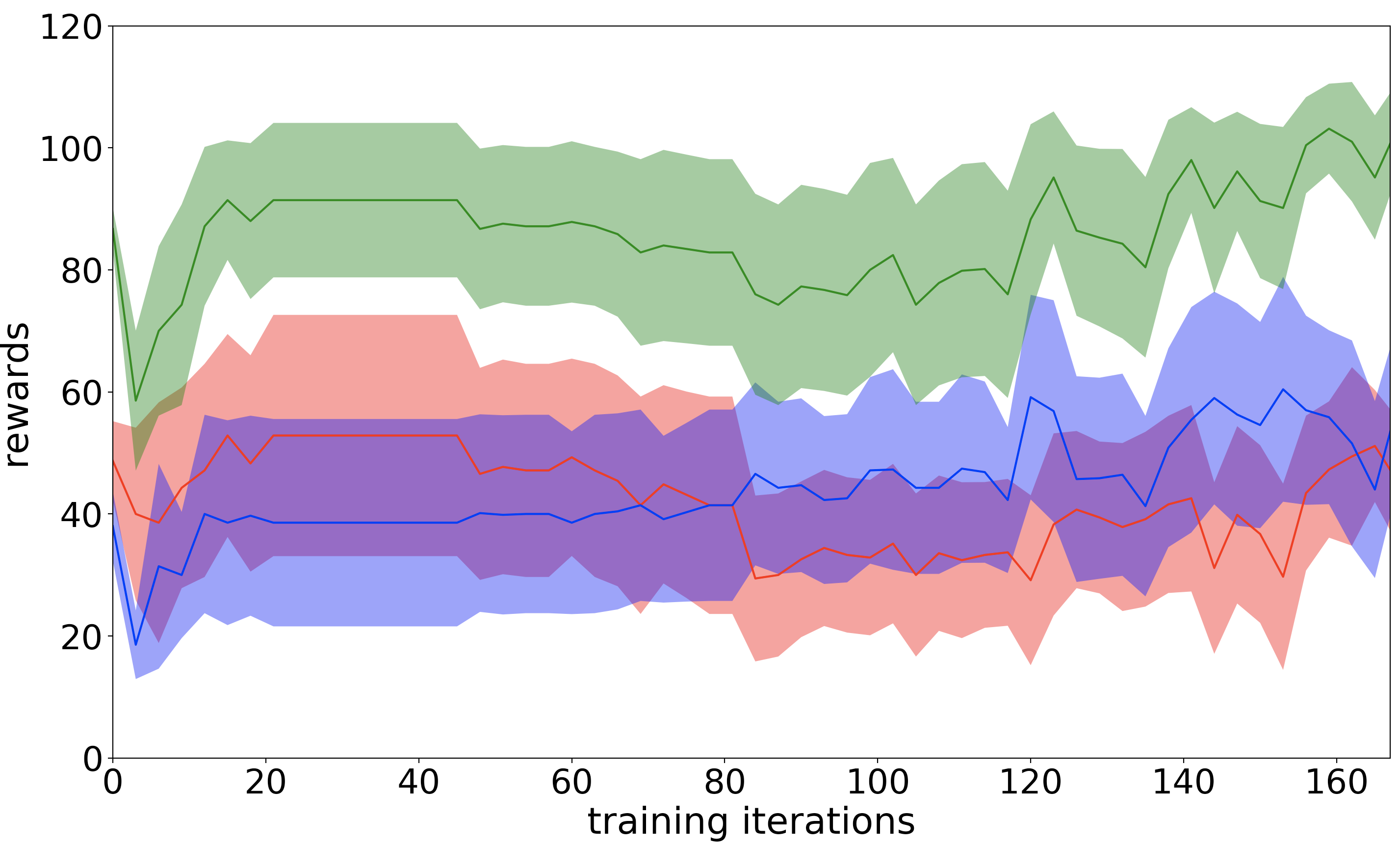}} \quad
  \subfigure[1 Punisher agent, 2 TFT agents]{\includegraphics[scale=0.11]{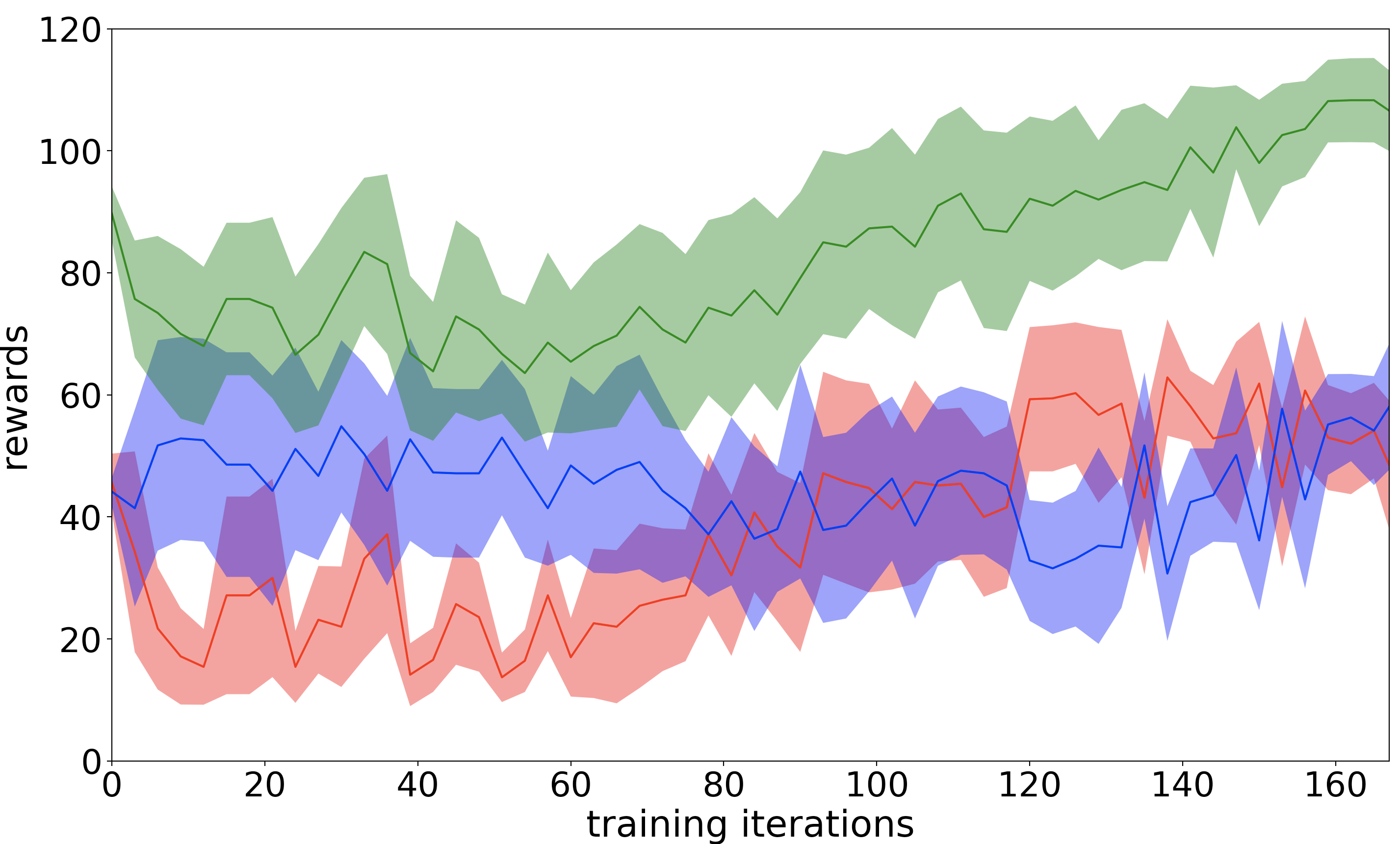}} \quad

  \label{multipleTFT}
  \caption{Adapting surroundings using TFT agents to get two independent Q-learning agents to cooperate in the IPD. (a) Q-learning agents learn independently to play the IPD and both learn defecting policies to achieve the minimum cumulative reward. As more TFT agents are introduced, cumulative reward rises. In (b) one TFT agent agents perform significantly better and in (c) two TFT agents both Q-learning agents learn to cooperate consistently achieving the maximum cumulative reward with less variance. We further display the changes in agent reward with a Sneaky agent and TFT agents, displayed in (d-f), and a Punisher agent and TFT agents, displayed in (g-i).}
\end{figure*}

\subsection{External Factors Influencing Agent Learning} 

The emphasis in current multiagent RL is to stationarize the environment using techniques that give the agent more information about the dynamics of the environment. However, in the previous section, we use an adaptation to training without changing the RL objective function or using a centralized control function. In this experiment we will also adapt training of Q-learning agents so they develop cooperative tendencies by inserting other agents into the environment that act as regulators. In open environments such as artificial societies, agents are likely to come into contact with unseen scenarios and their learning in these new environment is hard to predict. We are interested in understanding what happens when an agent is inserted in an unseen multi-agent environment with given dynamics, like a society with established social norms and how we can control for certain types of behavior without specifying changes to the agent's objective function. Evaluating these developmental aspects may provide key insights to understanding how types of behaviors are established in a society or how certain behaviors might provide the basis for stable social norms while others might not. We look to demonstrate in a simple, yet insightful, way that agents' behaviors may change in society based on their encounters with others and how this analysis can be useful for understanding social interactions. \\

We start with an environment that features only two players, both RL agents that train with a Q-learning update. TFT agents are added one-by-one to the environment to observe the changes in Q-values. Sneaky agents and Punisher agents are also added to see how cumulative reward increases and decreases. The format was modeled as a random encounters where each agent was matched with a random opponent with equal probability. Each match lasted for 20 timesteps, with each agent performing a total of 500 updates, one after each match. RL agents implemented $\epsilon$-greedy policies with $\epsilon$ decreasing linearly.

\section{Results and Discussion}

\subsection{Optimal Policies under Stationary Distributions} 
Individually, Q-learning agents and Hyper-Q learning agents score well against Axelrod agents due to their patterned behavior. The strategies that Axelrod agents use can be considered stationary and it is straightforward for the RL agents to learn an appropriate policy to play when matched against these agents provided that they have access to enough information to determine their strategy. In the case of the Sneaky agent, it requires at least four previous interactions at every time step in order to learn the optimal policy as it can then determine what the pattern in its behavior is. Both Q-learning and Hyper-Q-learning perform well against these agents. However, against the Prober agent, Hyper-Q performs significantly better and more consistently than either of the other two RL agents demonstrating an advantage of agent-tracking techniques in quasi-stationary environments. The average scores over 100 timesteps are recorded in Table \ref{Table: Static Agents}. Overall, when matched with the Axelrod agents, LTP agents achieve desirable results versus all the opponents and perform at least as well as Q-learning though Hyper-Q outperforms both. 

\subsection{Incorporating Information about Future Behaviors} In the case of LTP versus LTP agents, we observe that the agents are able to consistently achieve maximum cumulative reward when playing in the IPD. A variety of learned policies can be observed by setting $\eta$ to different values. By setting the value of $\eta$ to 0.99 (the same for all the agents), LTP agents heavily favor cooperation and consistently learn to cooperate with each other. In Figure \ref{Figure: avg RL rewards} we see the scores of LTP agents trained when $\eta = [0.01, 0.5, 0.7, 0.99]$. Between $\eta=0.01$ and $\eta=0.5$ there are no stark differences and both result in defecting policies. When $\eta=0.7$ there is a noticeable separation from a ``defect only''-type policy. LTP agents with $\eta$ values of less than 0.7 learn to defect following a similar trajectory to that of Q-learning and Hyper-Q-learning. With $\eta$ greater than $0.7$ both agents learn cooperative policies to achieve the maximum cumulative reward. When matching Q-learning agents with each other, they learn defecting policies every time over the course of 10,000 iterations. The number of times they defect increases exponentially until every action taken is to defect. While Hyper-Q-learning agents outperformed the other algorithms versus stationary agents, they performs poorly in this scenario. As the exploration rate decreases, the predicted probability of the opponent defecting increases and Hyper-Q agents learn to defect the quickest with negligible variance. We expect this will be further compounded using sampling techniques and buffer refreshing for social dilemmas as the new policies learned by its opponent will involve a higher likelihood to defect than before. Since the optimal policy to play against an opponent that is increasingly likely to defect is to defect oneself, predicting what the opponent might do next is not a viable strategy to maximize cumulative reward in the IPD. However, as LTP agents demonstrated, it is beneficial to take actions that maximize reward conditioned on changes in an opponent's behavior. 

\begin{table}
\begin{tabular}{c | c | c | c | c | c}
\centering
  & TFT & Punisher & Prober & Sneaky & FG \\
 \hline
 Q & 300 & 300 & 166.7  & 391.6 & 300 \\
 \hline
 Hyper-Q & 300 & 300 & 305 & 400 & 300 \\
\hline
LTP & 300 & 300 & 154.5 & 384.9 & 300 \\
\hline
\end{tabular}
\caption{Average scores vs. Static agents}
\label{Table: Static Agents}
\end{table}

\subsection{Cooperative Behavior in Presence of Stable Norms}
This experiment demonstrates how we can adjust the learning of our Q-learning agents by adapting the training procedure. We display the changes in the behavior of two Q-learning agents that would normally learn defect under a regular training procedure after interacting with regulating TFT agents. As can be seen in Figure \ref{Figure: avg RL rewards}, neither Q-learning agents nor agent-tracking agents like Hyper-Q learning learn to cooperate with one another when matched together in the IPD. However, we can show that tailoring their overall experience without explicitly changing the cost function or reward function, they can learn to cooperate with other RL agents which is an interesting find as it better represents how these agents would act in societal-like contexts rather than closed environments. By adding TFT agents to the environment and modeling it as random-encounters, the Q-values associated with cooperative behavior increase as the punishment for defecting is more likely to be immediate. In this way, the strategy of defecting is regulated by other existing agents. When facing other agents that are also exploring the environment, a Q-learning agent may be able to reap rewards from defecting behavior before other agents have adapted. This would cause the rest of the agents to also learn defecting behavior as shown in Figure \ref{Figure: avg RL rewards}. However, in a more strict environment, this behavior can be punished. By inserting more TFT agents into the environment, this behavior can be eliminated altogether and agents will learn to cooperate with one another consistently as shown in Figure 4. In addition, we can see that with only additional TFT agents in the environment, both agents achieve similar end rewards with neither consistently exploiting the other. In contrast, when a Sneaky agent is present, agent 1 barely manages to attain any reward as it has learned some cooperative behavior while agent 2 has learned primarily to defect due to the Sneaky agent's presence. When two TFT agents are introduced as shown in Figure 4 (f), defecting behavior can be regulated. However, one agent still manages to take advantage of the other and they do not achieve maximum cumulative reward as consistently. Surprisingly, the addition of a Punisher agent is not effective in producing cooperative behavior between agents. Though agents are punished in a similar fashion to TFT (immediate retaliation) this scenario lacks the consistency of TFT and the involvement of another RL agent exploring simultaneously. From these experiments we see that TFT is most successful out of the selected agents to regulate behavior and encourage RL agents to be cooperative. Hyper-Q agents do not learn to cooperate with each other in this setting. When paired versus other Hyper-Q learning agents, they continue to learn defecting policies when matched together and are unable to achieve the same results. In this regard, it is more difficult to regulate their strategies and are therefore these agents are less suitable for cooperation games. These initial experiments provide insights about the emergence of cooperation (or other behaviors) in the presence of environments or societies with stable pre-established dynamics, i.e., social norms.



\section{Conclusion}
In this paper we have presented a novel architecture for agents to learn optimal strategies for the IPD where elements of cooperation and competition are prevalent and important. Our LTP agents successfully learn to cooperate with one another by demonstrating changes to behavior via the use of probes and adjusting experiences to reflect these changes. We also show that these agents are able to learn optimal strategies when matched against stationary and quasi-stationary agents that have been used in Axelrod tournaments without adapting the objective function, focusing only on variations to training. We plan to focus on scaling this research to investigate more advanced social dilemmas and to incorporate the useful aspects of agent-tracking techniques to broaden the applicability of our approach. Building on this work, we also plan to study further how different types of behaviors may emerge in agent societies and how they might develop according to their surroundings. \\

\newpage


\newpage 
\bibliographystyle{ACM-Reference-Format}  
\bibliography{sample-bibliography}  

\end{document}